\documentclass[preprint,showpacs,prX]{revtex4}

\input epsf
\usepackage{graphicx}
\usepackage{epsfig}
\usepackage{mathrsfs}
\usepackage{amsfonts}
\usepackage{mathrsfs}
\usepackage{amsmath}
\usepackage{natbib}
\usepackage{epstopdf}
\usepackage{subfigure}

\begin{document}

\title{Shape Selection and Multi-stability in Helical Ribbons}

\author{Q. Guo$^{1,2*}$, A.K. Mehta$^{3}$, M.A. Grover$^{4}$, W. Chen$^{1,5}$, D.G. Lynn$^{3}$, Z. Chen$^{6*}$}
\affiliation{$^1$College of Materials Science and Engineering, Fuzhou University,Fujian 350108, China\\
$^2$Department of Mathematics and Physics, FuJian University of Technology,Fujian 350108, China\\
$^3$Departments of Chemistry and Biology, Emory University, Atlanta, Georgia 30322, USA\\
$^4$Department of Chemical and Biomolecular Engineering, Georgia Institute of Technology, Atlanta, Georgia 30332, USA\\
$^5$FuJian University of Technology, Fujian 350108, China\\
$^6$Department of Biomedical Engineering, Washington University, St. Louis, MO 63130 USA\\
* denotes equal contribution.
}

\date{\today}

\begin{abstract}
Helical structures, ubiquitous in nature, have inspired design and manufacturing of helical devices with applications in nanoelecromechanical systems, morphing structures, optoelectronics, micro-robotics and drug delivery devices. Meanwhile, multi-stable structures have attracted increasing attention for their applications in bio-inspired robots and deployable aerospace components. Here we show that mechanical anisotropy and geometric nonlinearity can lead to novel selection principle of shape and multi-stability in helical ribbons, with table-top experiments performed to demonstrate the working principle. Our work will promote understanding of large deformation and instability of thin objects, and serve as a tool in developing functional structures for broad applications.
\end{abstract}


\maketitle

\newcommand{\dx}{\mbox{${\bf d}_x$}}
\newcommand{\dy}{\mbox{${\bf d}_y$}}
\newcommand{\dz}{\mbox{${\bf d}_z$}}
\newcommand{\Ex}{\mbox{${\bf E}_x$}}
\newcommand{\Ey}{\mbox{${\bf E}_y$}}
\newcommand{\Ez}{\mbox{${\bf E}_z$}}
\newcommand{\dux}{\mbox{${\bf u}_1$}}
\newcommand{\duy}{\mbox{${\bf u}_2$}}
\newcommand{\drx}{\mbox{${\bf r}_1$}}
\newcommand{\dry}{\mbox{${\bf r}_2$}}
\newcommand{\strain}{\mbox{\boldmath $\gamma$}}
\newcommand{\andsp}{\mbox{$\quad\textrm{and}\quad$}}
\newcommand{\pfrac}[2]{\frac{\partial #1}{\partial #2}}


Many natural and synthetic materials exhibit helical shapes\cite{Helices_PNAS2006}, often driven by anisotropic pre-strains \cite{ArmonEfrati-116, Chen_APL2011, Gerbode_2012, Chen_2013}, swelling\cite{Mei_2012}, plastic coiling \cite{Seffen_2011}, mechanical instability \cite{Huang_SoftMatter2012}, lattice mismatch\cite{Zhang_nanolett2006, Dai_2013}, piezoelectricity\cite{kong_nano2003, MajidiChen-214}, molecular tilt\cite{Selinger_1996}, electrostatic interactions \cite{Terech_2012}, or differential growth \cite{Goriely_PRL1998, Wang_SR2013}. 
Helical structures also play a key role in engineering programmable matter that exhibits a variety of geometries and functionality in a delicately controlled fashion, and in such emerging stimulated by their applications in nano-elecromechanical systems (NEMS) \citep{Bunch_Science2007}, active materials \cite{Ge_APL2013}, drug delivery\cite{Hamley_2013}, morphing structures in aerospace engineering \cite{Lachenal_2012}, optoelectronics \citep{hwang_ijo2008}, and microrobotics \citep{abbott_ijrr2009}. 

Typically, helical ribbon shapes are achieved due to the balancing of surface stress or internal residual stress with elastic restoring forces of bending and stretching. The sophisticated interactions between elastic restoring forces and the molecular orientations and chirality often lead to the selection of different shapes\cite{Selinger_PRL2004,Sawa_PNAS2011, Sawa_2013}, such as cylindrical helical ribbons and tubules with vanishing Gauss curvature, and twisted ribbons or straight helicoids with non-zero Gauss curvature. For example, charged gemini surfactants with chiral counterions exhibit a transition between helical ribbons with cylindrical curvature and twisted ribbons with Gauss curvature as a function of molecular chain length\cite{Oda_1999}. 
Similarly, 
peptides organized as bilayer membranes \cite{Childers_2010} form ribbons that helically coil into micron-long nanotubes\cite{Lu_2003, Dong_2006, Childers_Langmuir2012}, and can transition reversibly between a purely twisted shape and a nanotube upon healing and cooling \cite{Hamley_2013}.

Driven by mechanical anisotropy (such as in surface stress or elastic modulus \cite{Wang_APL2008}) and geometric mis-orientation, helical shapes form spontaneously, and the transitions between helicoids, cylindrical helical ribbons, general helical ribbons and tubules can be achieved by tuning few geometric parameters such as the principal curvatures and mis-orientation angle\cite{Chen_APL2011, Chen_2013}. In twist-nematic-elastomer films, for example, the chiral molecular arrangement of liquid crystal mesogens drives the shape selection of helicoids and spiral ribbons due to the coupling between the order of liquid crystalline and elasticity\cite{Sawa_PNAS2011}. Nevertheless, the role of geometric nonlinearity in shape selection of helical geometries remains incompletely understood. In particular, geometric nonlinearity has recently been shown to be key in multi-stable structures\cite{Chen_PRL2012} featuring more than one stable shape arise in a variety of natural and engineering systems\cite{Forterre_Nature2005, Hyer_1981, Kebadze_2004, Daynes_2010, Vidoli_2008, Chen_2014}. Such structures have inspired design principles of deployable or smart actuation devices with multiple stable shapes each functioning in its own regime.  However, to our knowledge there have been few, if any, reports on multi-stable helical ribbon structures where both mechanical anisotropy and geometric nonlinearity are operating, although rod-like bistable helices have been studied through theory and experiments. Nor has the principle of shape selection and the nonlinear geometric effects in the formation of helical ribbons been well addressed.

In this letter, we show that the cooperation and compromise between the mechanical anisotropy and geometric nonlinearity lead to novel shape selection and multi-stability in spontaneous bending and twisting of ribbons.  A new theoretical framework has been proposed to address the origin of bistability due to geometric nonlinearity in square or rectangular plates, inspired by the bistable structures in nature and engineering, such as the Venus flytrap \cite{Forterre_Nature2005} and slap bracelets \cite{Kebadze_2004, Chen_2014}. Here we extend such work to study spontaneous bending and twisting of slender ribbons with mechanical anisotropy and geometric mis-orientation, and illustrate the principles of shape selection and instabilities of helical ribbons. Specifically, the transition from helical ribbons with non-zero Gauss curvature to nearly cylindrical helical ribbons with vanishing Gauss curvature occurs when the dimensionless parameter $\zeta$ goes beyond certain threshold. Moreover, the purely twisted ribbons can bifurcate into two nearly cylindrical helical ribbons with the same chirality but different bounding axes that are perpendicular to each other.

Our theory reveals the selection principles of shapes and multi-stability in helical geometries based on three-dimensional elasticity theory that incorporates mechanical anisotropy, geometric mis-orientatoin and nonlinear geometric effects. The shapes and spatial orientations of the helical geometries, the latter of which were often overlooked, are predicted quantitatively besides other geometric parameters (helix angle, radius and chirality), and validated through simple, table-top experiments. Our work will foster understanding of the morphogenesis and instability of thin objects in nature and engineering, and has important implications in designing new functional structures and devices that can be applied for optoelectronics \cite{hwang_ijo2008}, microrobots \cite{abbott_ijrr2009}, morphing structures\cite{Lachenal_2012}, and drug delivery systems\cite{Hamley_2013}.

\begin{figure}[t]
\begin{center}
\includegraphics[width=6in]{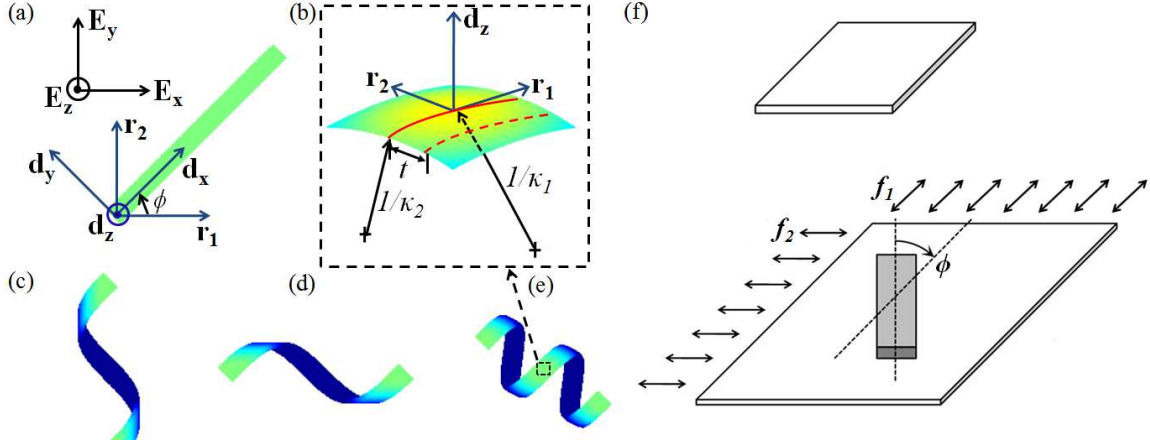}
\end{center}
\caption{Illustration of formation of a helical ribbon.  (a) The vectors $\dx$ and $\dy$ are oriented along the length and widthwise axes of the ribbon, respectively.  The bases ${\bf r}_1$ and ${\bf r}_2$ correspond to the principal axes of curvature. (b) Close-up view of a part of as-deformed helical ribbon in (e). (c) A cylindrical helical ribbon with $\kappa_2 = 0$. (d) A cylindrical helical ribbon with $\kappa_1 = 0$. (e) A general helical ribbon with non-zero $\kappa_1$ and $\kappa_2$. (f) Fabrication of a helical ribbon.  An elastic, adhesive strip was bonded to one pre-stretched sheet (or two sheets) of latex rubber, with a mis-orientation angle of $\phi$. }
\label{fig:helix}
\end{figure}

In our theoretical framework, the ribbon is modeled as an elastic strip with length $L$, width $w \ll L$, and thickness $H \ll w$ \cite{Chen_2013}.  The cross-section is rectangular and the principal geometric axes are along its length ($\dx$), width ($\dy$), and thickness ($\dz$). The originally flat ribbon lying along $\dx = \cos\phi \Ex + \sin\phi\Ey$ direction in the global Cartesian coordinate system. If the ribbon only bends along one principal axes (either ${\bf r}_1$ or ${\bf r}_2$) due to some surface or residual stresses, then a cylindrical helical ribbon with zero Gauss curvature forms as illustrated in Figure 1.(c) or (d) respectively. A general helical ribbon (with moderately small width), however, can bear principal curvatures $\kappa_1$ and $\kappa_2$ are along the axes ${\bf r}_1$ and ${\bf r}_2$ with a mis-orientation angle $\phi$ within the ribbon plane\cite{Chen_APL2011,Chen_2013}. Then a point ${\bf P}$ (${\bf P}(s) = X(s)\Ex + Y(s)\Ey + Z(s)\Ez$) on the centerline of the deformed ribbon can be parameterized by the arclength $s$:$X(s) = s - (\beta^2/\alpha^3) (\alpha s - \sin \alpha s)$, $Y(s) = (\beta \tau/\alpha^3) (\alpha s - \sin \alpha s)$, and $Z(s) =  (\beta/\alpha^2)(\cos \alpha s - 1)$,  where $C \equiv \cos \phi$, $S \equiv \sin \phi$, $\alpha = \sqrt{\kappa_1^2 C^2 + \kappa_2^2 S^2}$, $\beta = \kappa_1 C^2 + \kappa_2 S^2$, and $\tau = (\kappa_1 - \kappa_2)S C$ (for detailed derivations, see Ref. \cite{Chen_2013}).

The geometric parameters of the helical ribbon can be determined from the values of $\kappa_1$, $\kappa_2$, and $\phi$. For example, the helix angle between the central axis of the bounding cylinder and the widthwise axis ($\dy$) of the ribbon is $\Phi = \arctan { \{(\kappa_1 - \kappa_2) S C/ (\kappa_1 C^2 + \kappa_2 S^2) \}}$. The radius of the bounding cylinder is $R = (\kappa_1 C^2 + \kappa_2 S^2)/(\kappa_1^2 C^2 + \kappa_2^2 S^2)$. In addition, the helix chirality is set by the sign of the helix angle, $ \mathop{\mathrm{sgn}}(\Phi)$. It is also worth noting that not only are these geometric parameters are derived, but the orientation of the helix can be determined, e.g., the helix axis is ${\bf M} =  \frac{\tau}{\alpha} \Ex + \frac{\beta}{\alpha} \Ey$. The orientation can be important in some engineering applications, for example, the anomalous coiling of helical semiconductor nanoribbons \cite{Zhang_nanolett2006} featuring variations of helix pitch, angle, radius and orientation.

\begin{figure}[t]
\begin{center}
\includegraphics[width=4in]{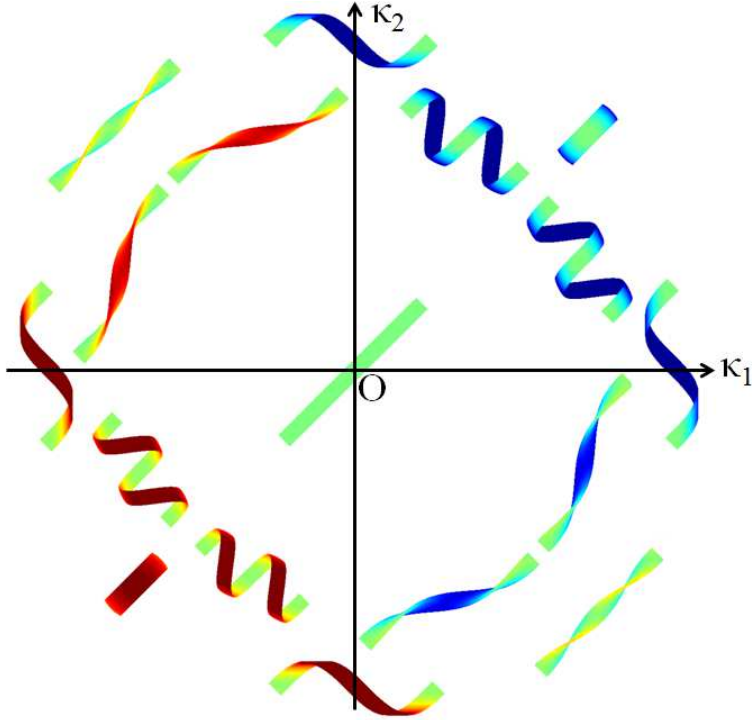}
\end{center}
\caption{Versatile ribbon morphology and orientation controlled by the geometric parameters $\kappa_1$, $\kappa_2$ and $\phi$. Here $\phi = \pi/4$ is fixed, and the ribbon morphology and orientation vary as the relative values of $\kappa_1$ and $\kappa_2$ vary.}
\label{fig:Kinematics_study}
\end{figure}

A variety of shapes including helical cylindrical shapes, rings, purely twisted ribbons, with different spatial orientations as shown in Figure \ref{fig:Kinematics_study}, can be achieved just by tuning the values of $\kappa_1$, $\kappa_2$ (or equivalently the mean curvature and Gauss curvature) and mis-orientation angle $\phi$ (which is nevertheless fixed in Figure \ref{fig:Kinematics_study}). Although geometrically distinct, each shape represents an element in a subset of a complete class of two-dimensional manifolds controlled by these three independent geometric variables\cite{Chen_2013}.

To obtain the values of $\kappa_1$, $\kappa_2$ and $\phi$, we employ a theoretical model based on linear elasticity theory, differential geometry and stationarity principles, which takes into consideration both the non-uniform bending and mid-plane stretching due to geometric nonlinearity. We consider the conformation of a small piece of the ribbon onto the surface of a torus to account for the geometric nonlinearity. The total potential energy density per unit area of the ribbon is $\Pi = {\bf f}^-:\strain |_{z=-H/2} + {\bf f}^+:\strain |_{z=H/2} + \int_{-H/2}^{H/2}\frac{1}{2}\strain:{\bf C}:\strain\,dz $, where ${\bf C}$ is the fourth-order elastic constant tensor.  It is worth noting that for an elastically anisotropic ribbon, the principal bases of ${\bf C}$ may not coincide with $\{\dx,\dy,\dz\}$.  At equilibrium, $\Pi$ must be stationary with respect to the unknown parameters $\kappa_1$ and $\kappa_2$, i.e.,  $\partial\Pi/\partial\kappa_i = 0$ ($i = 1,2$).

We have previously shown\cite{Chen_APL2011,Chen_2013} that it is sufficient to consider only the deformation when the ribbon is subjected to an effective surface stress on the bottom surface ($z = -H/2$), ${\bf f}^* = f_1 \dux \otimes \duy  + f_2 \dux \otimes \duy$, because the problem with two surface stress tensors acting on both top and bottom surfaces can be reduced to a problem with only one surface stress, when only the bending mode is of practical interest for the mechanical self-assembly of helical ribbons. To arrive at concise, analytic solutions, we assume the material properties are isotropic and homogeneous, with Young's modulus $E$ and Poisson's ratio $\nu$. But in principle, the theoretical framework adopted herein can be generalized to study more complex systems with heterogeneity.

For the isotropic case under consideration, the potential energy per unit length of the strip can be more explicitly written as $\Pi \approx - (f_1 \kappa_1 + f_2 \kappa_2)WH/2 + A (\kappa_1^2 + \kappa_2^2 + 2\nu \kappa_1 \kappa_2) EWH^3 + B EHW^5 (\kappa_1 \kappa_2)^2$, where $A = 1/24(1-\nu^2)$ and $B = 1/[640(1-\nu^2)] $. Here, $\Pi_b \sim (\kappa_1^2 + \kappa_2^2 + 2\nu \kappa_1 \kappa_2) EWH^3$ denotes the energy penalty due to bending, and $\Pi_g \sim EHW^5 (\kappa_1 \kappa_2)^2$ is the extra stretching energy due to geometric nonlinearity. When the dimensionless geometric parameter equals its threshold value (i.e., $\eta \equiv W \sqrt{\kappa/H} = \eta_c = [80(1+\nu)/3]^{1/4}$ \cite{Chen_PRL2012, ArmonEfrati-116}, a characteristic width is obtained such that $W_c = [80(1+\nu)/3]^{1/4}\sqrt{H/\kappa} \approx 2.5 \sqrt{H/\kappa}$ (where $\kappa \equiv \texttt{max}\{|\kappa_1|, |\kappa_2|\}$). It is worth noting that $\eta$ involves the unknown parameter $\kappa$, so it is more convenient to use an equivalent dimensionless parameter, $\zeta \equiv \sqrt{f/EH} W/H$ \cite{Chen_PRL2012}, for the design purpose, where $f \equiv \texttt{max}\{|f_1|, |f_2|\}$. In this case, $W_0 = \zeta_0 H \sqrt{EH/f}$.

In the small width regime ($W \ll W_0$), or equivalently, when ($\zeta \ll \zeta_0$), applying stationarity principles yields $\kappa_1 = 6(f_1 - \nu f_2)/E H^2$, $\kappa_2 = 6(f_2 - \nu f_1)/E H^2$. 
This corresponds to the cases discussed in the previous studies by Chen et al.\cite{Chen_APL2011, Chen_2013}. In the large width regime ($W \gg W_0$ or $\zeta \gg \zeta_0$), by contrast, the geometric nonlinearity requires that either $\kappa_1 = 0$ or $\kappa_2 = 0$ for the most parts (except near the edges), otherwise the stretching energy $\Pi_g$ is going to be very large compared with the bending energy $\Pi_b$, and hence inadmissible. This leads to bifurcated solutions $\phi = 0$ and $\kappa_1 = 6(1 - \nu^2)f_1/E H^2$ and $\phi = \pi/2$ and $\kappa_1 = 6(1 - \nu^2)f_2/E H^2$, corresponding to the potential energy $\Pi^*(\phi) = - 3(1 - \nu^2) f_i^2 W/2 E H$ ($i = 1,2$) respectively. These results are consistent with the Stoney's formula with a modified Young's modulus\cite{Stoney_1909}. Obviously, bending along the large principal surface stress direction is more energetically favorable. It can also be shown that the ribbon is bistable only when $f_g \equiv f_1 f_2 < 0$. The corresponding examples are illustrated in the following experiments (Fig. \ref{fig:experiment3A}).


\begin{figure}[t]
\begin{center}
\includegraphics[width=4in]{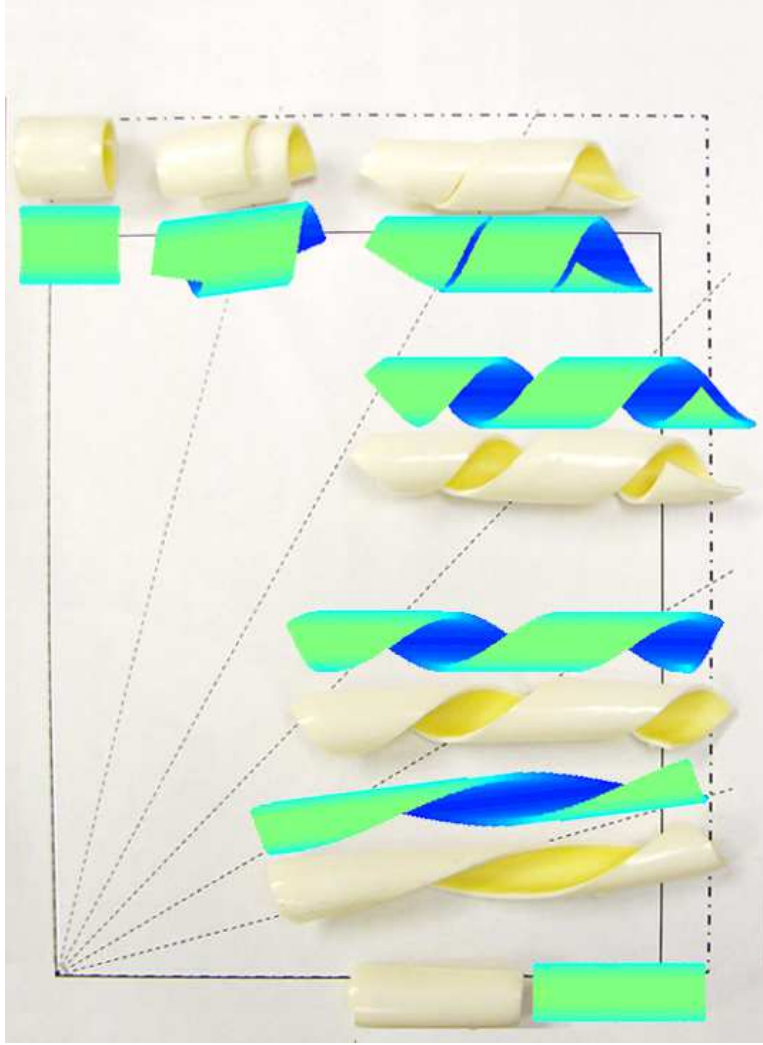}
\end{center}
\caption{Mono-stable helical ribbons with nearly zero Gauss curvature.  A strip of elastic, pressure sensitive adhesive was bonded to a pre-stretched sheet of latex rubber, with a mis-orientation angle $\phi$ going from $0$ to $\pi/2$ at a $\pi/12$ interval. The pre-stretches are $q = 0.24$ and $p = 0.12$ in the vertical and horizontal directions, respectively. The color is indexed according to the out-of-plane displacement to help visualize the 3D deformation as in the following figures. Yellow denotes zero displacement. A darker color indicates a larger out-of-plane displacement.}
\label{fig:experiment1}
\end{figure}

Our theoretical framework predicts shapes selection of ribbons of tunable morphologies and associated multi-stability. To compare with the theoretical predictions, a series of simple, table-top experiments were performed to achieve a variety of shapes and different multi-stability.  A sheet (or two sheets) of latex rubber was pre-stretched and bonded to an elastic strip of thick, pressure-sensitive adhesive~\cite{Chen_2013}. The pre-stretches in the principal directions are $p$ and $q$ respectively. For convenience, we define ``Gauss surface stress'' as $f_g = (f_1 - \nu f_2) (f_2 - \nu f_1)$, since this quantity will dictate the design of Gauss curvature ($\kappa_G = \kappa_1\kappa_2 = 36(f_1 - \nu f_2)(f_2 - \nu f_1)/E^2 H^4$). Here, $f_1 = pE_1h$ and $f_2 = qE_1h$ are the principal components of the effective surface stress, where $h = 0.3$mm is the thickness of the latex sheet, and $E_1$ is its Young's modulus. In the scenario where ``Gauss surface stress'' is positive ($f_g > 0$), it was previously observed that helical ribbons with non-zero Gauss curvature, varying radius and bounding axes arose in the small width cases when $W = 12.5$mm $< W_0 \approx 15.4$mm (as shown in Figure 2 of Ref. \cite{Chen_APL2011}). Drastically different from these observations, the new helical ribbons appear with nearly zero Gauss curvature, almost identical helix radius and bounding axis when the width of the ribbon goes well beyond the critical width for multi-stability ($W = 24.0$mm $\gg W_0 \approx 15.4$mm), even though the effective surface stresses stay the same (Figure \ref{fig:experiment1}). The theoretical predictions of the ribbon shapes are in good agreement with the experimental results.

\begin{figure}[h!]
\begin{center}
\includegraphics[width=4in]{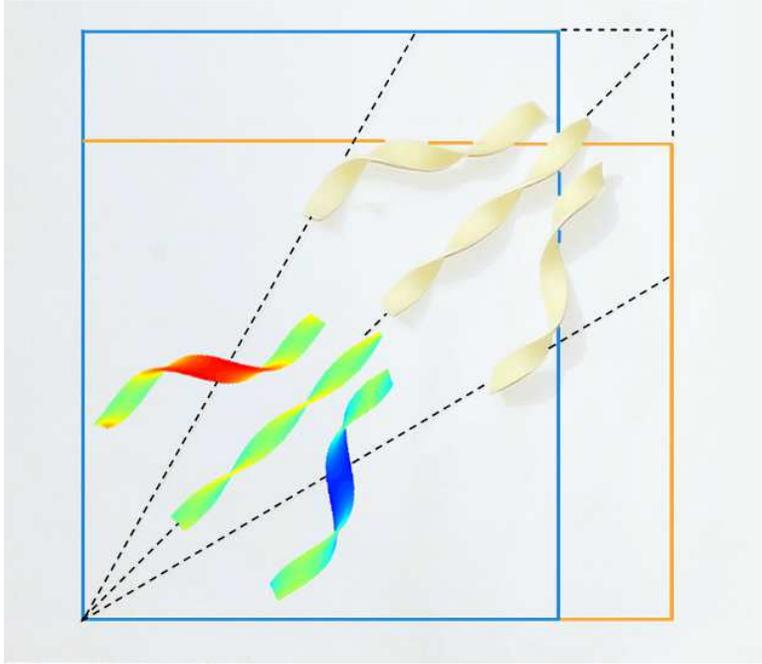}
\end{center}
\caption{Mono-stable helical ribbons with negative Gauss curvatures.  A narrow elastic strip was bonded to two sheets of latex rubber pre-stretched in perpendicular directions, with a mis-orientation angle of $\phi = \pi/6$, $\pi/4$, and $\pi/3$. The pre-stretches in the top latex sheet are $p = 0, q = 0.24$, and those in the bottom sheet are  $p = 0.24, q = 0$.}
\label{fig:experiment2}
\end{figure}

\begin{figure}[h!]
\begin{center}
\includegraphics[width=3.5in]{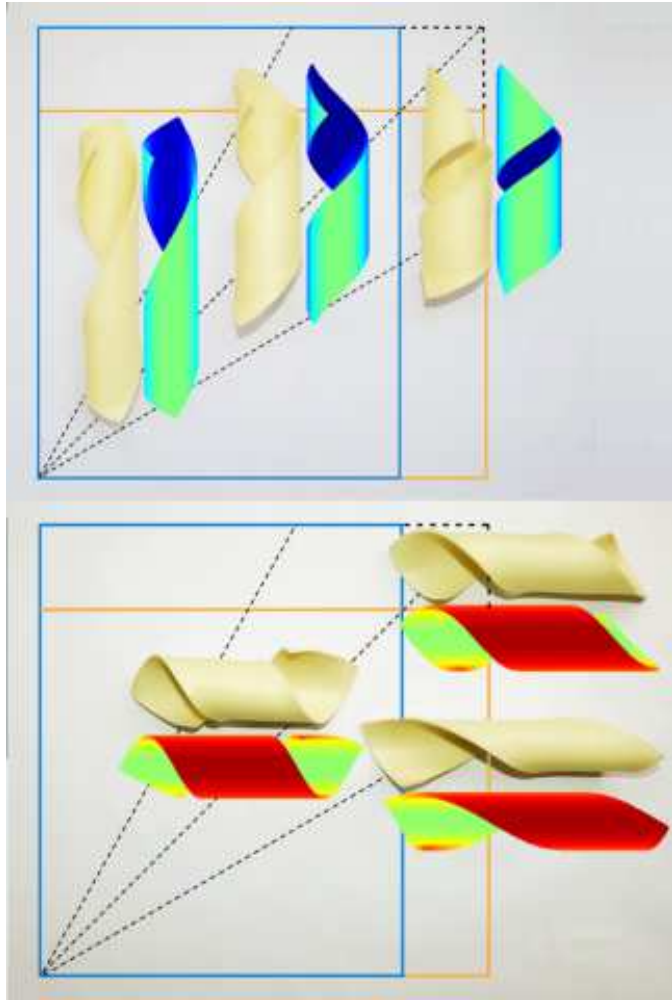}
\end{center}
\caption{Bistable helical ribbons with nearly zero Gauss curvatures.  A wide elastic strip  was bonded to two sheets of latex rubber pre-stretched in perpendicular directions, with a mis-orientation angle of $\phi = \pi/6$, $\pi/4$, and $\pi/3$. The bonded multi-layer system bifurcated into either one of the two nearly cylindrical shapes. The pre-stretches in the top latex sheet are $p = 0, q = 0.24$, and those in the bottom sheet are  $p = 0.24, q = 0$. The color is indexed according to the out-of-plane displacement. Yellow denotes zero displacement. Blue indicates the displacement is negative (below the paper plane), while red indicates a positive displacement (above the paper plane).}
\label{fig:experiment3A}
\end{figure}

For ribbons subjected to a negative ``Gauss surface stress'' ($f_g < 0$), however, a bifurcation occurs when the width increases and goes beyond the threshold, leading to the dramatic change from a mono-stable helical ribbon with negative Gauss curvature to one of the two nearly cylindrical helical ribbons with almost zero Gauss curvature. Figure \ref{fig:experiment2} shows good agreements between theoretical predictions and the experiments when the width is smaller than the critical width ( when $W = 12.5$mm $< W_0 \approx 15.4$mm). In this case, mono-stable helical ribbons with negative Gauss curvature form. Subjected to the same set of surface stresses, however, the helical ribbons would spontaneously deform into one of the two nearly cylindrical shapes when the width went well beyond the threshold (i.e., $W = 48.0$mm $\gg W_0 \approx 15.4$mm), as shown in Figure \ref{fig:experiment3A}. For example, a purely twisted ribbon with zero bounding radius and $\pi/2$ helix angle formed when $f_1 = -f_2$, $\phi = \pi/4$ and the width was small enough (see Figure \ref{fig:experiment2}). However, as the width became much larger than the threshold, the same set of surface stresses resulted in either one of the nearly cylindrical helical ribbons with non-zero bounding radius and $\pi/4$ helix angle. The transitions between one locally state (with a helix angle of around 60 degrees) and the other (with a helix angle of around 30 degrees) are demonstrated in the Supplemental Movie.

\begin{figure}[h!]
\begin{center}
\includegraphics[width=4in]{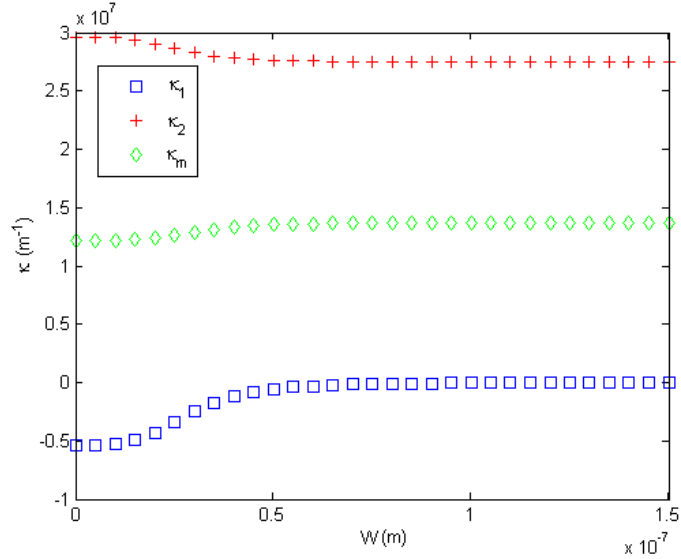}
\end{center}
\caption{The equilibrium values of the principal and mean curvatures for ribbons of different widths ($\nu = 0.5$).}
\label{fig:TheoPCurv}
\end{figure}

\begin{figure}[h!]
\begin{center}
\includegraphics[width=4in]{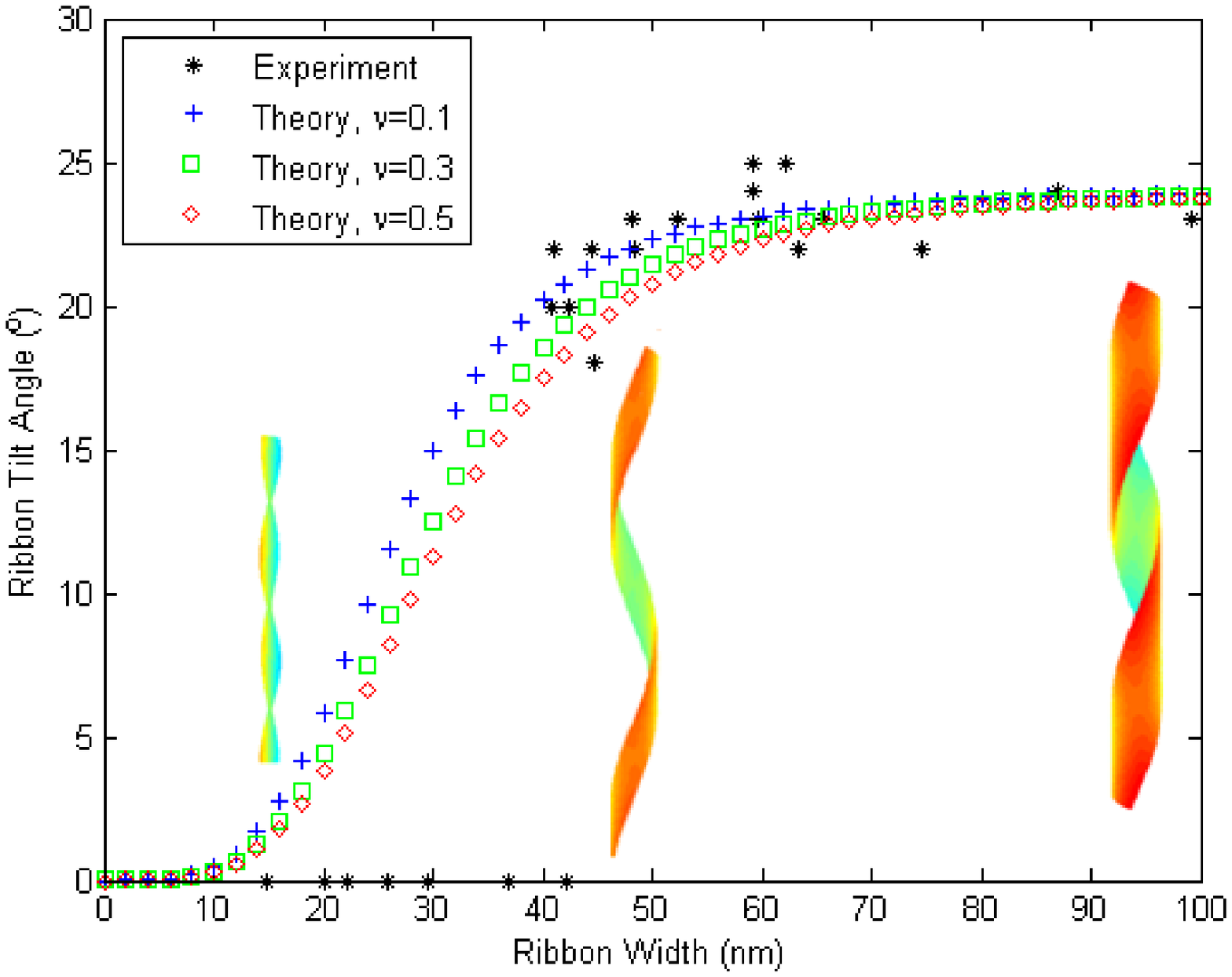}
\end{center}
\caption{Comparisons between theoretical predictions and experimental observations (black dots)\cite{Childers_Langmuir2012} of the tilt angle as a function of the ribbon width. Geometrically, the aggregates change morphology from a purely twisted shape (when the width is very small) with zero tilt angle, to a helical ribbon with an intermediate tilt angle,  to a cylindrical helical ribbon of a nearly constant tilt angle of around $24^{o}$. Three different values of the Poisson's ratio ($\nu = 0.1$, $0.3$, and $0.5$) are used in the calculations to show the influence of the Poisson's ratio on the sharpness of the shape transition.}
\label{fig:TheoryVsExp}
\end{figure}

We also applied our theory to the shape transitions in the central hydrophobic nucleating core of the A$\beta$ peptide of Alzheimer's disease, A$\beta$(16-22) \cite{Lu_2003}. In solution, this peptide assembles into purely twisted ribbons or cylindrical helical ribbons \cite{Childers_Langmuir2012}. In the case of purely twisted ribbons \cite{Childers_Langmuir2012}, we notice that the tilt angle is zero when the ribbon is very narrow, which indicates that it is a purely twisted ribbon such that $\kappa_1^{0} C^2 + \kappa_2^{0} S^2 = 0$, where $\kappa_1^{0} = 6(f_1 - \nu f_2)/E H^2$, $\kappa_2^{0} = 6(f_2 - \nu f_1)/E H^2$ are the principal curvatures in the limit that the width goes to zero. On the other hand, in the large width regime, the tilt angle approaches a nearly constant value of $24^{o}$, featuring a nearly cylindrical helical ribbon with an helix angle $67^{o}$, thickness $H = 4.3$nm and radius of curvature $R^{inf} = 26$nm. This indicates that $\kappa_1^{\inf} = 0$ and $\kappa_2^{\inf} = 1/R^{inf} = 6(1 - \nu^2)f_2/E H^2$.
By numerically solving the set of linearly independent equations, $\partial \Pi/\partial \kappa_1 = \partial \Pi/\partial \kappa_2 = 0$, we obtained the following equilibrium values of the principal curvatures for ribbons of different widths (see Fig. \ref{fig:TheoPCurv}). Here, since that the Poisson's ratio $\nu$ is currently unavailable, we have used three different values ($\nu = 0.1$, $0.3$, and $0.5$) in calculating the principal curvatures. Correspondingly, the tilt angle varies as a function of the ribbon width, because of the change in the relative values of the principal curvatures. Fig. \ref{fig:TheoryVsExp} shows comparisons between theoretical predictions and experimental observations of the tilt angle as a function of the ribbon width. The trend of the theoretically predicted curve with a given Poisson's ratio captures the main asymptotic features of the experimental data. The theoretical curve predicts a gradual increase, with the derivative first increasing then decreasing over a relatively wide domain, in contrast to the sharp transition at around 40nm observed in experiments. Indeed, as the Poisson's ratio is decreased, the transition to a helically coiled ribbon becomes sharper. These results suggest that this theory is not completely capturing the molecular cooperativity present in these peptide bilayers assemblies, which awaits further investigation.

In this work, shape selection of helical ribbons and their multi-stability are shown to result from co-functioning of mechanical anisotropy, geometric mis-orientation and geometric nonlinearity. The novel selection principles are established through a comprehensive, three-dimensional theory that combines linear elasticity, differential geometry and stationarity principles. The chirality of the helical ribbon does not change when it goes from one stable state to the other. The quantitative relationship between the surface stress, elastic moduli, helix angle, radius and chirality is established.  When subjected to arbitrary effective surface stresses, an initially flat, straight ribbon would deform either into a helical form with non-zero Gauss curvature throughout, or a nearly cylindrical shape with zero Gauss curvature for the main interior regions. When the ``Gauss surface stress'' is positive (i.e. the principal surface stresses have the same sign), the nearly cylindrical shape bends long the axis on which the principal curvature has a higher magnitude where the total potential energy is locally and globally minimized. While if the ``Gauss surface stress'' is negative (i.e.the principal surface stresses have different signs), the system bifurcates into one of the two locally stable states, bending about one of the two principal axes.

This work generalizes the results of our previous work on helical ribbons with relatively small width that bear non-zero Gauss curvatures when subject to non-zero ``Gauss surface stresses''.  Table-top experiments were also performed with composites formed by bonding a pre-stretched sheet (or two sheets) of elastomer with an elastic layer of soft acrylic, demonstrating the design principle of tunable geometries with desirable mechanical multi-stability.  Our work has important implications on the abnormal coiling shapes of helical semiconductor, peptides, and nematic ribbons, where the variations in dimensions (i.e., the width and thickness) lead to helical ribbons of different shapes and orientations. A comprehensive theoretical framework is established for predicting and designing the shapes of helical structures with tunable parameters and predictable instability that can be used in a broad range of biological and engineering applications. The current results can also be pertinent to the spontaneous curling, twisting, buckling and wrinkling of graphene sheets \cite{Wang_2012} and strained multilayer structures\cite{Suo_APL1999, Zhang_nanolett2006, Dai_2013} where both geometric nonlinearity and mechanical anisotropy play important roles on the geometric shape, mechanical, electronic and optical properties of materials.

{\it Acknowledgements} -- The authors thank Drs. M. Haataja, D. Srolovitz, C. Majidi and H. Stone for helpful discussions and comments. This work has been in part supported by the National Science Foundation of China Grant No.11102040, the Foundation of Fujian Educational Committee (Grant No.JA12238), and the Sigma Xi Grants-in-Aid of Research (GIAR) program. Z.C. acknowledges support from the Society in Science - Branco Weiss fellowship, administered by ETH Z$\ddot{u}$rich. The authors (DGL) acknowledge the Division of Chemical Sciences, Geosciences, and Biosciences, Office of Basic Energy Sciences of the U.S.Department of Energy through Grant DE-ER15377 for peptide synthesis and analyses and the NASA Astrobiology Program, under the NSF Center for Chemical Evolution, CHE-1004570 and NSF-CBC-0739189 for support.


\begin{thebibliography}{00}


\bibitem{Helices_PNAS2006} N. Chouaieb, A. Goriely, and J. H. Maddocks. \textbf{Proc. Natl. Acad. Sci. USA} \textbf{103}, 9398 (2006).

\bibitem{ArmonEfrati-116}S. Armon, E. Efrati, R. Kupferman, and E. Sharon, \textbf{Science} \textbf{333}, 1726-30 (2011).

\bibitem{Chen_APL2011} Z. Chen, C. Majidi, D.J. Srolovitz, and M. Haataja \textbf{Appl. Phys. Lett.} \textbf{98}, 011906 (2011).

\bibitem{Gerbode_2012} S.J. Gerbode, J.R. Puzey, A.G. McCormick, L. Mahadevan.  {\bf Science} {\bf 337}, 1087-1091 (2012).

\bibitem{Chen_2013} Z. Chen, C. Majidi, D.J. Srolovitz, and M. Haataja, arXiv: 1209.3321, submitted to \textbf{Proc. R. Soc. A}.


\bibitem{Mei_2012} W. Li, G. Huang, H. Yan, J. Wang, Y. Yu, X. Hu, X. Wu, and Y. Mei, 
Soft Matter {\bf{8}}, 7103-7107 (2012).

\bibitem{Seffen_2011} Seffen KA, Guest SD, Prestressed morphing bistable and neutrally stable shells. \textbf{J. Appl. Mech.} \textbf{78}, 011002 (2011).

\bibitem{Huang_SoftMatter2012} J. Huang, J. Liu, B. Kroll, K. Bertoldi, and D.R. Clarke, \textbf{Soft Matter} \textbf{8}, 6291-6300 (2012).

\bibitem{Zhang_nanolett2006} L. Zhang, E. Ruh, D. Gr\"utzmacher, L. Dong, D.J. Bell, B.J. Nelson, and C. \& Sch\"onenberger, \textbf{Nano Lett.} \textbf{6}, 1311-1317 (2006).


\bibitem{Dai_2013} L. Dai, and L. Zhang. \textbf{Nanoscale} \textbf{5}, 971-976 (2014).


\bibitem{kong_nano2003} X.Y. Kong and Z.L. Wang, \textbf{Nano Lett.} \textbf{3}, 1625 (2003).

\bibitem{MajidiChen-214} C. Majidi, Z. Chen, D.J. Srolovitz, and M. Haataja, \textbf{ J. Mech. Phys. Solids} \textbf{58}, 73-85 (2010).

\bibitem{Selinger_1996} J.V. Selinger, F.C. MacKintosh, and J.M. Schnur, \textbf{Phys. Rev. E} \textbf{53}, 3804 (1996).

\bibitem{Terech_2012} P. Terech, S.K.P. Velu, P. Pernot, and L. Wiegart, \textbf{J. Phys. Chem. B} \textbf{116}, 11344-11355 (2012).

\bibitem{Goriely_PRL1998} A. Goriely, and M. Tabor, \textbf{Phys. Rev. Lett.}  \textbf{80}, 1564-1568 (1998).

\bibitem{Wang_SR2013} J. Wang, G. Wang, X. Feng, T. Kitamura, Y. Kang, S. Yu, and Q. Qin, \textbf{Scientific Reports} 3, 3102 (2013).

\bibitem{Bunch_Science2007} J.S. Bunch, A.M. Van Der Zande, S.S. Verbridge, I.W. Frank, D.M. Tanenbaum, J.M. Parpia, H.G. Craighead, and P.L. McEuen, \textbf{Science} {\bf{315}}, 490-493 (2007). 

\bibitem{Ge_APL2013} Q. Ge, H.J. Qi, and M.L. Dunn. \textbf{Appl. Phys. Lett.} \textbf{103}, 131901 (2013).

\bibitem{Hamley_2013} I.W. Hamley, A. Dehsorkhi, V. Castelletto, S. Furzeland, D. Atkins, J. Seitsonen, and J. Ruokolainen, \textbf{Soft Matter} \textbf{9}, 9290-93 (2013).





\bibitem{Lachenal_2012} X. Lachenal, P.M. Weaver, and S. Daynes, \textbf{Proc. R. Soc. A} \textbf{468}, 1230-51 (2012).

\bibitem{hwang_ijo2008} G. Hwang, C. Dockendorf, D. Bell, L. Dong, H. Hashimoto, D. Poulikakos, and B. Nelson, \textbf{Int. J. Optomechatroni.} \textbf{2}, 88 (2008).


\bibitem{abbott_ijrr2009} J.J. Abbott, K.E. Peyer, M.C. Lagomarsino, L. Zhang, L.X. Dong, I.K. Kaliakatsos, and B.J. Nelson, \textbf{Int. J. Rob. Res.} \textbf{28}, 1434 (2009).

\bibitem{Sawa_PNAS2011} Y. Sawa, F. Ye, K. Urayama, T. Takigawa, V. Gimenez-Pinto, R.L. Selinger, and J.V. Selinger \textbf{Proc. Nat. Acad. Sci. USA} \textbf{108},6364-6368 (2011).

\bibitem{Sawa_2013} Y. Sawa, K. Urayama, T. Takigawa, V. Gimenez-Pinto, B.L. Mbanga, F. Ye, J.V. Selinger, and R.L. Selinger, \textbf{Phys. Rev. E} \textbf{88}, 022502 (2013).

\bibitem{Selinger_PRL2004} R.L.B. Selinger, J.V. Selinger,  A.P. Malanoski, and J.M. Schnur, \textbf{Phys. Rev. Lett.} \textbf{93}, 158103 (2004).

\bibitem{Oda_1999} R. Oda, I. Huc, M. Schmutz, S.J. Candau, and F.C. MacKintosh, \textbf{Nature} \textbf{399}, 566-569 (1999).

\bibitem{Childers_2010} W.S. Childers, A.K. Mehta, R. Ni, J.V. Taylor, D.G. Lynn DG, \textbf{Angew Chem, Int Ed.} \textbf{49}, 4104-4107 (2010).

\bibitem{Lu_2003} K. Lu, J. Jacob, P. Thiyagarajan, V.P. Conticello, and D.G. Lynn. \textbf{J. Am. Chem. Soc.} \textbf{125}, 6391-3 (2003).

\bibitem{Dong_2006} J. Dong, K. Lu, A. Lakdawala, A.K. Mehta, and D.G. Lynn, \textbf{Amyloid} \textbf{13}, 206-215 (2006).



\bibitem{Childers_Langmuir2012} W.S. Childers, N.R. Anthony, A.K. Mehta, K.M. Berland, and D.G. Lynn, \textbf{Langmuir}, \textbf{28}, 6386-6395 (2012).


\bibitem{Wang_APL2008} J. Wang, X. Feng, G. Wang, and S. Yu, \textbf{Appl. Phys. Lett.} \textbf{92}, 191901 (2008).


\bibitem{Chen_PRL2012} Z. Chen, Q. Guo, C. Majidi, W. Chen, D.J. Srolovitz, and M.P. Haataja, \textbf{Phys. Rev. Lett.} \textbf{109}, 114302 (2012).


\bibitem{Forterre_Nature2005} Y. Forterre, J.M. Skotheim, J. Dumais and L. Mahadevan, \textbf{Nature}, \textbf{433}, 421-25 (2005).

\bibitem{Hyer_1981} M.W. Hyer, \textbf{J. Compos. Mater} \textbf{15}, 296-310 (1981).

\bibitem{Kebadze_2004} E. Kebadze, S.D. Guest, and S. Pellegrino, \textbf{Int. J. Solids Struct.} \textbf{41}, 2801¨C2820 (2004).

\bibitem{Daynes_2010} S. Daynes, C.G. Diaconu, K.D. Potter, and P.M. Weaver, \textbf{J. Compos. Mater} \textbf{44}, 1119-1137 (2010).

\bibitem{Vidoli_2008} S. Vidoli, and C. Maurini, \textbf{Proc. R. Soc. A} \textbf{464},  2949-2966 (2008).

\bibitem{Chen_2014} Q. Guo, H. Zheng, W. Chen, and Z. Chen, \textbf{Biomed. Mater. Eng.} \textbf{24}, 557-562 (2014).




















\bibitem{Stoney_1909} G.G. Stoney, \textbf{Proc. R. Soc. A} \textbf{82}, 172 (1909).


\bibitem{Wang_2012} H. Wang, and M. Upmanyu, \textbf{Nanoscale} \textbf{4}, 3620-3624 (2012).

\bibitem{Suo_APL1999} Z. Suo, E.Y. Ma, H. Gleskova, and S. Wagner, \textbf{Appl. Phys. Lett.} \textbf{74}, 1177 (1999).

















\end{thebibliography}
\end{document}